\documentclass[12pt]{article}
\usepackage{latexsym} 
\usepackage{a4}
\usepackage{amsfonts} 
\newcommand{\Mvariable}{\alpha}
\newcommand{\Res}[1]{{\raisebox{-2mm}{\rm Res}\atop { #1}}}

\newcommand{\Ref}[1]{(\ref{#1})}
\newcommand{\pa}{\partial}\newcommand{\al}{\alpha}
\newcommand{\be}{\begin{equation}}\newcommand{\ee}{\end{equation}}
\newcommand{\bea}{\begin{eqnarray}}\newcommand{\eea}{\end{eqnarray}}
\newcommand{\beao}{\begin{eqnarray*}}\newcommand{\eeao}{\end{eqnarray*}}
\newcommand{\M}{{\cal M}}

\begin{document}
\begin{center}
{\bf \Large Heat Kernel Expansion for Semitransparent Boundaries}
\end{center}
\vspace{12pt}
\begin{center}
M. Bordag\footnote{e-mail: Michael.Bordag@itp.uni-leipzig.de}
and D.V. Vassilevich\footnote{On leave from Department of Theoretical Physics,
     St.Petersburg University,
     198904 St.Petersburg, 
     Russia. e-mail: Dmitri.Vassilevich@itp.uni-leipzig.de}\\[5pt]
{\small Institute for Theoretical Physics, Leipzig
University\\Augustusplatz 10/11, 04109 Leipzig, Germany}
\begin{abstract}
We study the heat kernel for an operator of Laplace type with a
$\delta$-function potential concentrated on a closed surface. We
derive the general form of the small $t$ asymptotics and calculate
explicitly several first heat kernel coefficients.
\end{abstract}
\end{center}
\section{Introduction}
Singular potentials are a frequently used idealization of physical
situations allowing for an easier (and, sometimes, explicit) solution
while keeping the essential features of the problem.  The best studied
cases of singular potentials are the delta function potentials
concentrated at isolated points, which describe contact interactions
of particles (for a review, see \cite{Albeverio}). Rigorous analysis
of such potentials was initiated by the paper by Berezin and Faddeev
\cite{BerezinFaddeev} and has developed later in a mature mathematical
discipline \cite{Albeverio}.  Other cases of singular potentials
include also cosmic strings and other topological defects
\cite{Vilenkin} and problems related to the black hole entropy
\cite{Frolov:1998vs}.  We like also to mention a recent work on the
boundary discontinuities \cite{Apps:1997zr}.

With respect to the Casimir effect, a delta function shaped potential
provides the simplest generalization of the conductor boundary
conditions towards inclusion of more realistic properties of the walls
like partial transparency. For a scalar and a spinor field with plane
boundaries this problem has been investigated in \cite{bhr}, for
moving partly transmitting mirrors in \cite{reyn}. An interesting
approach using 'semihard' and 'weak' boundaries is developed in
\cite{actor}.  In all those cases it is crucial to know the ultra
violet divergences in order to find out the structure of necessary
counterterms.  This is equivalent to the investigation of the
corresponding heat kernel asymptotics. As it is known \cite{Gilkey},
this is an expansion with respect to integer powers of the proper time
parameter $t$ (see below) for Laplace type operators on closed
manifolds and to half integer powers, i.e., to powers of $\sqrt{t}$,
on manifolds with boundaries and Dirichlet or Robin boundary
conditions.  For more complicated pseudo differential operators powers
of $\ln t$ may appear.

The present paper is devoted to singularities which are located on
closed hypersurfaces of dimension $m-1$ where $m$ is dimension of the
underlying manifold. Apart from the quantum mechanical problem of a
particle in the space with semitransparent boundaries (for a recent
calculation of the vacuum energy for such system see
e.g. \cite{scandurra}) possible physical application also include
fermions on a background of a magnetic tube \cite{Bordag:1998tg} and
photons interacting with dielectric bodies \cite{Schwinger,bkv99}. In
all these cases the dynamics is described by a second order
differential operator of Laplace type supplemented by certain matching
conditions on a surface. Our primary interest is in the heat kernel
asymptotics. They govern the short time asymptotics of quantum
mechanical propagators, the ultra violet divergences and the large
mass expansion in quantum field theory.

The heat kernel coefficients for singular potentials cannot be
obtained in general as limiting cases of smooth configurations.  This
is clear already from the fact that the heat kernel expansion for a
smooth potential contains powers of the potential taken at the same
point. Such expressions become ill-defined in the delta-function
limit.  In our previous paper \cite{bkv99} we have observed a
surprising property that sometimes the ultra violet behavior of a
system with $\delta$-function potential is less singular than that of
corresponding ``smooth'' system.

Let us proceed with basic definitions.  Let $M$ be a smooth Riemannian
manifold of dimension $m$.  Let $\Sigma$ be a smooth closed
submanifold of co-dimension $1$. Let ${\cal V}$ be a vector bundle
over $M$.  Let $E$ and $V$ be endomorphisms of ${\cal V}$ and ${\cal
V}|_{\Sigma}$ respectively.  In more ``physical'' notations, $E$ and
$V$ are matrix valued functions bearing spin and internal indices.  In
the present paper we study the heat kernel expansion for the operator
\be D =-(\nabla^2 +E(x) +\delta_\Sigma V(x)) \label{P} =D_0
-\delta_\Sigma V(x) \,.  \ee Let $dx$ and $dy$ be the Riemannian
volume elements on $M$ and $\Sigma$.  We normalize the function
$\delta_\Sigma$ in such a way that for any smooth function $f$ \be
\int_M \delta_\Sigma f(x) dx= \int_\Sigma f(y) dy \,.  \ee We adopt
the following short hand notations for the integrals \be \int_M dx\,
F(x)=\{ F\} [M] \,, \qquad \int_\Sigma dy\, F(y)=\{ F\} [\Sigma ] \,.
\ee

We can choose the coordinates in such a way that in the
vicinity of $\Sigma$ the metric has the form
\be
g_{ij}dx^idx^j =(dx^m)^2 + g_{ab}dx^adx^b \,.
\label{metric}
\ee
The second fundamental form of $\Sigma$ is 
$L_{ab}=\frac 12 \partial_m g_{ab}$.
We suppose that $x^m=0$ on $\Sigma$.
Our notations are the same as in Ref.\cite{a5bgv}. $R_{ijkl}$
are the components of the Riemann curvature tensor. With our
sign conventions, $R_{1212}$ is negative on the standard
sphere in Euclidean space. The Ricci tensor
$  \rho $ and the scalar curvature
$  \tau $ are given by
\be
\rho  \sb{ ij}:=R \sb{ ik kj} \qquad
 \tau = \rho  \sb{ ii}=R \sb{ ik
ki}.
\nonumber
\ee
Let
$  \rho  \sp{ 2}:= \rho  \sb{ ij} \rho  \sb{
ij}$ and
$ R \sp{ 2}:=R \sb{ ij kl}R \sb{ ij kl}$
be the norm of the Ricci and full curvature
tensors. Let $\Omega_{ij}$ be the endomorphism valued
components of the curvature of the connection on ${\cal V}$.
In physical language, $\Omega_{ij}$ is the field strength for
the Yang-Mills and spin-connections. 
Let \lq ;\rq\ denote multiple covariant differentiation
with respect to the Levi-Civita connection
of
$  M$ and let \lq:\rq\ denote multiple
tangential covariant differentiation on the
boundary with respect to the Levi-Civita
connection of
$  \Sigma ;$ the difference between these
two is measured by the second fundamental
form. For example, $f_{;jj}=f_{;mm}+f_{;aa}=
f_{;mm}+f_{:aa}-L_{aa}f_{;m}$.

A mathematically rigorous way to define the spectral problem
for the operator $D$ is to replace it by the spectral problem
for $D_0$ for $x\not\in \Sigma$ supplemented by the conditions
on $\Sigma$ (see e.g. \cite{Albeverio})
\begin{eqnarray}
&&\phi (-0)=\phi (+0) \label{cond1} \\
&&\nabla_m \phi (-0)-\nabla_m\phi (+0)=V\phi(0)
\label{cond2}
\end{eqnarray}
with the short hand notation $\phi (\pm 0)=\lim_{x^m\to\pm 0}\phi (x)$.

The heat kernel $K(x,y;t)$ is a solution of the heat equation
\be
(\partial_t +D_0) K(x,y;t)=0 \label{hkeq}
\ee
with the initial condition
\be
K(x,y;0)=\delta (x,y) \,,\label{incond}
\ee
which satisfies the matching conditions on $\Sigma$ following
from (\ref{cond1}) and (\ref{cond2}).

We are interested in the integrated heat kernel
\be
K(f,D;t)=\{ f(x)K(x,x;t) \}[M]={\rm Tr} (f\exp (-tD))\,.
\label{inthk}
\ee
On manifolds with boundary with local boundary conditions
there is an asymptotic expansion as $t\to +0$
\be
K(f,D;t)=\sum_{n=0} a_n (f,D)t^{(n-m)/2}\,,
\label{asymptotex}
\ee where the coefficients $a_n(f,D)$ are volume and surface integrals
of local invariants. The existence of the asymptotic expansion
(\ref{asymptotex}) is usually considered as granted. There are however
important exceptions (besides ``genuine'' pseudo differential
operators, the square root of Laplacian for instance), such as the
boundary value problem for spectral boundary conditions and
$\delta$-function potentials with point-like support on manifolds with
dimension $m\ge 2$.  In such cases $\ln t$ terms can appear in the
asymptotic expansion \cite{Grubb,albeverio2,Solodukhin}.  To the best
of our knowledge, the existence of the expansion (\ref{asymptotex})
for the problem considered here has never been stated before.

This paper is organized as follows. In the next section we derive an
integral equation for the heat kernel and show validity of the
asymptotic expansion (\ref{asymptotex}).  In section 3 we calculate
the heat kernel asymptotics for the particular case when $\Sigma$ is a
sphere in ${\bf \rm R}^m$. In section 5 we derive explicit expressions
for the heat kernel coefficients $a_n(f,D)$, $n\le 5$ for the most
general form of the operator $D$. To this end we use the particular
case calculations of section 4 and functorial properties of the heat
kernel.

\section{General structure of the heat kernel}
To study the general structure of the heat kernel expansion
we use an integral equation similar to that proposed
by Gaveau and Schulman \cite{Gaveau} for the one-dimensional
$\delta$-potential:
\be
K(x,y;t)=K_0(x,y;t) +\int_0^t ds \int_{\Sigma}dz\,
K_0(x,z;t-s)V(z)K(z,y;s) \,
\label{inteq}
\ee
where $K_0(x,y;t)$ denotes the heat kernel corresponding
to the operator $D_0$ with $V=0$.

The equation (\ref{inteq}) has a solution in the form
of the power series in $V$:
\begin{eqnarray}
K(x,y;t)=K_0(x,y;t) + \sum_{n=1}^\infty
\int_0^t ds_n \int_0^{s_n}ds_{n-1} \dots \int_0^{s_2}ds_1
\int_\Sigma dz_n \dots \int_\Sigma dz_1 \nonumber
&& \\
 \quad \times K_0(x,z_n;t-s_n)V(z_n)K_0(z_n,z_{n-1};s_n-s_{n-1})
\dots V(z_1)K_0(z_1,y;s_1) \label{solint} \,.&&
\end{eqnarray}
The equation (\ref{inteq})
can be obtained formally as a limiting case of the smooth potential.
Instead of investigating this limiting procedure we prefer 
to check directly that the
heat kernel defined by (\ref{inteq}) satisfies the heat equation
(\ref{hkeq}) with the initial condition
(\ref{incond}) and boundary conditions which follow from
(\ref{cond1}) and (\ref{cond2}). The initial condition
(\ref{incond}) is evident from the eq. (\ref{solint}).
Only the first term contributes at $t=0$ if $x,y\not\in\Sigma$.
The heat equation (\ref{hkeq}) can be checked by a direct
calculation. The first of the matching conditions (\ref{cond1})
just expresses the fact that the heat kernel $K_0(x,y;t)$
is smooth enough. The second condition (\ref{cond2}) is a bit
less trivial. Let $y\not\in\Sigma$. Then the following sequence of
transformations holds:
\begin{eqnarray}
&&\nabla^x_m K(-0,y;t)-\nabla^x_mK(+0,y;t)=\nonumber \\
&&\qquad =-\lim_{\epsilon\to 0} \int_{-\epsilon}^\epsilon dx^m 
\nabla^2_m \int_0^t ds \int_{\Sigma}dz\,
K_0(x,z;t-s)V(z)K(z,y;s) \nonumber \\
&&\qquad =\lim_{\epsilon\to 0} \int_{-\epsilon}^\epsilon dx^m
D_0^x \int_0^t ds \int_{\Sigma}dz\,
K_0(x,z;t-s)V(z)K(z,y;s) \nonumber \\
&&\qquad =\lim_{\epsilon\to 0} \int_{-\epsilon}^\epsilon dx^m
\int_0^t ds \int_{\Sigma}dz\,
(\partial_s K_0(x,z;t-s))V(z)K(z,y;s) \nonumber \\
&&\qquad =\lim_{\epsilon\to 0} \int_{-\epsilon}^\epsilon dx^m
\int_0^t ds \int_{\Sigma}dz\,
K_0(x,z;t-s)V(z)D_0^yK(z,y;s) \nonumber \\
&&\qquad +\lim_{\epsilon\to 0} \int_{-\epsilon}^\epsilon dx^m
\int_{\Sigma}dz\, [\delta (x,z) V(x) K(z,y;t)-K_0(x,z;t)V(z)\delta (z,y)]
\nonumber \\
&&\qquad =V(x^a,0)K((x^a,0),y;t) \nonumber
\end{eqnarray}

The main advantage of the representation (\ref{solint}) is that
the small $t$ behavior of $K(x,y;t)$ is defined through the
small $t$ behavior of $K_0(x,y;t)$ which is known in some detail.
To simplify the notations we do not write down explicitly the
volume elements here and the parallel transport matrices later on.
We also drop all matrix indices.

We are interested in the integrated heat kernel $K(t)=\int dx K(x,x;t)$,
where we put the smearing function $f=1$ for simplicity. The integration
over $x$ can be performed by using the equation
\be
\int_M dx\, K_0 (x,y_1;\tau_1) K_0(y_2,x;\tau_2)=K_0(y_1,y_2;\tau_1+\tau_2)\,,
\label{prodK}
\ee
which follows from the evident operator identity 
$e^{-\tau_1D_0}e^{-\tau_2D_0}=e^{-(\tau_1+\tau_2)D_0} $. We have
\begin{eqnarray}
K(t)=K_0(t)+\sum_{n=1}^\infty
\int_0^t ds_n \int_0^{s_n}ds_{n-1} \dots \int_0^{s_2}ds_1
\int_\Sigma dz_n \dots \int_\Sigma dz_1 \nonumber
&&\\
 \qquad \times K_0(z_1,z_n;t+s_1-s_n)V(z_n)K_0(z_n,z_{n-1};s_n-s_{n-1})
\dots V(z_1)\,. &&\label{solK}
\end{eqnarray}

It is instructive to calculate explicitly
several first terms of the expansion (\ref{solK}). In the linear
order in $V$ one immediately gets
\be
K_1(t) =t\int_\Sigma dz K_0(z,z;t)V(z) \label{K1} \,.
\ee
This equation means that the liner order in $V$ can be
obtained from the asymptotic expansion for the heat kernel
in which a smooth potential is replaced by the singular one.
Such simple relation does not hold at higher orders in $V$.

In analyzing the order $V^2$ contributions we suppose that
$D_0$ is just the standard scalar Laplacian in ${\bf \rm R}^m$,
neglect derivatives of $V$ and suppose that $\Sigma$ is flat.
However, we turn to a bit more general case allowing $\Sigma$
to be of dimension $m-k$. In this particular case 
$K_0(x,y;t)=(4\pi t)^{-m/2} \exp (-(x-y)^2/4t)$ and
\be
K_2(t)=(4\pi )^{-\frac{m+k}2} t^{\frac{m-k}2}
\int_\Sigma dz\, V(z)^2 \int_0^t ds_2\int_0^{s_2} ds_1
\left( (t-s_2+s_1)(s_2+s_1) \right)^{-\frac k2} \,.
\label{K2k}
\ee
For $k>1$ the integrals over $s$ are divergent. This shows
that the proposed method can not be extended in particular to  
$\delta (r)$ potentials in ${\rm \bf R}^m$ with $m\ge 2$
for which the expansion (\ref{asymptotex}) is known to
break down \cite{albeverio2,Solodukhin}).
Return to the subject of our study, $k=1$. All integrals
are easily calculated giving
\be
K_2(t)=\frac{t}{(4\pi t)^{(m-1)/2}} \frac 18 \int_\Sigma dz V(z)^2
+\dots \,,
\label{K21}
\ee
where we have omitted all higher order terms.

To calculate the other terms in the expansion (\ref{solK}) one can use
the following strategy. Let us substitute the small $s$ asymptotic
expansion for the $K_0(x,y;s)$
\be
K_0(x,y;s)\sim \frac{\exp (-\sigma (x,y)/2s)}{(4\pi s)^{m/a}}
(A_0^{(0)}(x,y)+sA_2^{(0)}(x,y)+\dots ) \,,
\label{expK0}
\ee
where $\sigma (x,y)$ is the half square of the geodesic distance between
$x$ and $y$, $A_{2i}^{(0)}(x,y)$ are the heat kernel coefficients
for the operator $D_0$.
Contributions of largely separated points to (\ref{solK}) are 
exponentially damped. Therefore, we may expand 
$A_{2i}^{(0)}(z_j,z_{j-1})$ and $\sigma (z_j,z_{j-1})$
in Taylor series in $(z_j-z_{j-1})$. The potentials $V(z_j)$
are to be expanded around certain point, say $z_1$. At the end,
the expression under the integrals in (\ref{solK}) will become a sum
of the monomials
\begin{eqnarray}
&&\exp \left( -\frac{(z_1-z_n)^2}{4(t-s_1+s_n)} -\dots -
\frac{(z_2-z_1)^2}{4(s_2-s_1)} \right) I^{(N)} (V,\dots ;z_1)
\nonumber \\
&&\quad \times \frac{(z_1-z_n)^{N_1}}{(t-s_1+s_n)^{M_1}} \dots
\frac{(z_2-z_1)^{N_n}}{(s_2-s_1)^{M_n}} \,,
\label{mono}
\end{eqnarray}
where $I^{(N)}$ is a local invariant functional of $V$, geometric invariants
and their derivatives calculated at the point $z_1$. Negative powers
of $(s_j-s_{j-1})$ appear due to the expansion of the $\sigma (z_j-z_{j-1})$
in the exponentials. It is easy to see that $M_j\le N_j/2$.

Integrals over $z_i$, except for the last one over $z_1$,
can be calculated with the help of the relation
\be
\int_{{\rm \bf R}^n} d x\exp \left( -\frac {(y_1-x)^2}{4\alpha}
-\frac {(x-y_2)^2}{4\beta} \right) =
\left( \frac{4\pi \alpha\beta }{\alpha +\beta} \right)^{\frac n2}
\exp \left( -\frac{(y_1-y_2)^2}{4(\alpha +\beta )} \right)
\label{integral}
\ee
with positive real parameters $\alpha$ and $\beta$.
Note that eq. (\ref{integral}) is just a particular case of 
(\ref{prodK}) when $D_0$ is the flat space Laplacian. Integrals
of even powers of $(x-y_1)$ with the same exponential weight
are obtained by differentiation of (\ref{integral}) with respect
to $\alpha$. Odd powers of $x$ can be integrated by using the
following obvious relation
\be
\int_{{\rm \bf R}^n} d x\, \left( x^a -\frac{\beta y_1^a + \alpha y_2^a}{
\alpha +\beta} \right)
\exp \left( -\frac {(y_1-x)^2}{4\alpha}
-\frac {(x-y_2)^2}{4\beta} \right) =0 \,.\label{linint}
\ee
Before integrating over $s_j$ let us introduce rescaled variables
$\tilde s_j=s_j/t$. This enables us to extract an overall
power of the proper time $t$. Integrals over $\tilde s_j$
will give just numerical factors. One can easily see that
the strongest possible singularity of the integrand has the
form $((1-\tilde s_n +\tilde s_1)\dots (\tilde s_2 -\tilde s_1))^{-\frac 12}$.
This singularity is integrable. Hence the integral over $\tilde s_j$
always exists. After all the integrations have been done one
obtains $\int_\Sigma dz_1 I^{(N)}(V,\dots ;z_1)$ multiplied by
a numerical coefficient and a power of $\sqrt t$.

Generalization of the procedure proposed in this section for the
case of non-unit smearing function $f$ is obvious. Therefore,
we have demonstrated that for the spectral problem considered
in this paper the asymptotic expansion (\ref{asymptotex}) is valid
where the coefficients $a_n(f,D)$ are integrals over $M$ and $\Sigma$
of local invariants. Volume terms are the same as in the heat kernel
expansion for the operator $D_0$.

\section{Penetrable spherical shell}
Before calculating the heat kernel expansion for generic
form of the operator $D$ consider a particular case of
the constant potential $V$ with the support on a spherical
shell of the radius $R$:
\be
V(x) \delta_\Sigma =-\frac{\alpha}R  \delta(r-R) \,.\label{sphpot}
\ee
Here $r$ is the radial coordinate.
Let $D_0$ in (\ref{P}) be the standard scalar Laplacian in
${\bf \rm R}^m$.
After separating the angular variables
the eigenvalue equation takes the form
\be
\left(-{\pa^{2}\over \pa r^{2}}-{(m-1)\over r}{\pa\over\pa r}+{l(l+m-2)\over
    r^{2}} \right)\phi=k^{2}\phi \,,
\ee
where $l=0,1,2,\dots$ is the orbital momentum. After the
substitution $\phi(r)= r^{(2-m)/2} \psi(r)$ the equation takes the form
\be\label{eq2}
\left(-{\pa^{2}\over \pa r^{2}}-{1\over r}{\pa\over\pa r}+{\nu^{2}\over
    r^{2}} \right)\psi=k^{2}\psi \,,
\ee
where the notation $\nu=l+{m-2\over 2}$ is introduced.

Using the standard techniques \cite{standardtechniques}, 
the zeta function associated with this operator
 can be written in the form
 \be \zeta(s)=\int\limits_{0}^{\infty}~{{\rm d} t \over t}~{t^{s}\over \Gamma
   (s)}~ K(t)\,, \ee
where $K(t)$ is the integrated heat kernel. 
The function $\Gamma(s)\zeta(s)$ has simple poles in $s=\frac{d}{2}-N$
($N=0,\frac12,2,\dots$) whose residua are determined by the   expansion
of the heat kernel for $t\to 0$. It reads
\be
K(t)\sim \sum\limits_{n\ge 0}a_{n} t^{(n-m)/2}\,.
\ee
It can be easily seen that the coefficients are related to the residua by
means of
\be a_{n}=\Res{s=\frac{m-n}{2}}\Gamma(s)\,\zeta(s)\,.
\label{anRes}
\ee
These heat kernel coefficients
 can be obtained by calculating the zeta function starting
from the differential equation, Eq. \Ref{eq2}.  The zeta function of the
operator $D$ can be expressed in the form
\be\label{zeta2}
\zeta(s)={\sin \pi s\over \pi} \ \sum\limits_{l=0}^{\infty} \ D_{l} \
\int\limits_{0}^{\infty} {\rm d}k \ k^{-2s} {\pa\over\pa k} \ \ln f_{l}(ik)\,,
\ee
where $f_{l}(k)$ is the Jost function of the scattering problem corresponding
to the operator $D$ and
\be\label{mult}
D_{l}={(2l+m-2)~(l+m-3)!\over l!~(m-2)!}\qquad m\ge 3
\ee
is the multiplicity of the orbital eigenvalues.  This representation, as it
stands, is valid for $\Re s >\frac{m}{2}$. The Jost function reads
\cite{bkv99}
\be\label{jostfct}
f_{l}(ik)=1+\al I_{\nu}(k)K_{\nu}(k)\,.
\ee
In Eq. \Ref{zeta2} the contribution resulting from the empty space (it
is independent from $\alpha$) is dropped. By this reason there is no
contribution corresponding to the coefficient $a_{0}$ there.
 
In fact, we need the residua of the function
$\Gamma(s)\zeta(s)$. These are delivered when inserting the uniform
asymptotic expansion for $k\to\infty$ and $l\to\infty$ of the Jost
function into Eq. \Ref{zeta2}. The latter can be obtained simply by
inserting the known expansions of the Bessel functions into
\Ref{jostfct}. This expansion can be written in the form
\be\label{expansion}
\ln f_{l}(ik)=\sum\limits_{n,i} \ X_{n,i} \ t^{i} \ \nu^{-n}\,.
\ee
Here, $n=1,\dots,N_{m}$, and $N_{m}$ is the number of the highest heat
kernel coefficient requested to be calculated.  The coefficients
$X_{n.i}$ are numbers, some first given in the appendix.  The notation
$t=\sqrt{1+(k/\nu)^{2}}$ is used.  When inserting \Ref{expansion} into
the r.h.s. of Eq. \Ref{zeta2}, the change of variables $k\to \nu k$
can be made after what the expression factorizes. The integral can be
calculated easily. It reads
\be\label{int} 
\int\limits_{0}^{\infty} \ {\rm d}k \ k^{-2s} {\pa\over\pa k}
  t^{i} = - {\Gamma(1-s)\Gamma(s+i/2)\over\Gamma(i/2)}\,.  
\ee
The sum over $l$ takes the form 
\be\label{sum}
\zeta(d,n)=\sum\limits_{l=0}^{\infty} D_{l}\,\nu^{-2s-n}\,.
\ee
The functions $\zeta(d,n)$ can be expressed in terms of Riemann and Hurwitz
zeta functions. They are shown in the appendix.

Now the heat kernel coefficients can be expressed in the form

\be\label{hkks} a_{N}=-\Res{s=\frac{d-N}{2}}
\sum_{n=1}^{N_{m}}\zeta(d,n)\sum_{i}X_{n,i}{\Gamma(s+i/2)\over\Gamma(i/2)}
\quad (N=1,\frac32,2,\frac52,\dots, N_{m})\,.  
\ee
The sum over $i$ runs from $i=n$ to $i=n+4(\left[{n+1\over
2}\right]-1)$ where $[\dots]$ denotes the integer part. In this form
they can be calculated immediately using one of the standard computer
systems for analytical calculations.  As a results, we obtain that
$a_1=0$ in any dimension. Other coefficients read for $m=3$:
\begin{eqnarray}
&&a_{2}={\frac{-\Mvariable}{2\,{\sqrt{\pi }}}}\qquad
a_{3}={\frac{{{\Mvariable}^2}}{8}}\nonumber \\
&&a_{4}={\frac{-{{\Mvariable}^3}}{12\,{\sqrt{\pi }}}}\qquad
a_{5}={\frac{{{\Mvariable}^4}}{64}}\nonumber\\
&&a_{3}={\frac{-\left( {{\Mvariable}^3}\,
   \left( 4 + 21\,{{\Mvariable}^2} \right)  \right) }{2520\,
   {\sqrt{\pi }}}}\qquad
a_{7}={\frac{{{\Mvariable}^4}\,
   \left( 1 + 2\,{{\Mvariable}^2} \right) }{1536}}\,,
\label{m3}
\end{eqnarray}
for $m=4$:
\begin{eqnarray}
&&a_{2}={\frac{-\Mvariable}{8}}\qquad
a_{3}={\frac{{{\Mvariable}^2}\,{\sqrt{\pi }}}{32}}
\nonumber\\
&&a_{4}={\frac{-{{\Mvariable}^3}}{48}}\qquad
a_{5}={\frac{-\left( {{\Mvariable}^2}\,
     \left( 3 - 4\,{{\Mvariable}^2} \right) \,{\sqrt{\pi }} \right) }
   {1024}}\label{m4}\\
&&a_{6}={\frac{-\left( {{\Mvariable}^3}\,
    \left( -8 + 7\,{{\Mvariable}^2} \right)  \right) }{3360}}\qquad
a_{7}={\frac{-\left( {{\Mvariable}^2}\,
    \left( 135 + 168\,{{\Mvariable}^2} - 
      128\,{{\Mvariable}^4} \right) \,{\sqrt{\pi }} \right) }{3932\
  16}} \,,\nonumber
\end{eqnarray}
for $m=5$:
\begin{eqnarray}
&&a_{2}={\frac{-\Mvariable}{12\,{\sqrt{\pi }}}}\qquad
a_{3}={\frac{{{\Mvariable}^2}}{48}}\nonumber\\
&&a_{4}={\frac{-{{\Mvariable}^3}}{72\,{\sqrt{\pi }}}}\qquad
a_{5}={\frac{{{\Mvariable}^2}\,
     \left( -2 + {{\Mvariable}^2} \right) }{384}}\label{m5}\\
&&a_{6}={\frac{-\left( {{\Mvariable}^3}\,
       \left( -24 + 7\,{{\Mvariable}^2} \right)  \right) }{5040\,
    {\sqrt{\pi }}}}\qquad
a_{7}={\frac{-9\,{{\Mvariable}^4} + 2\,{{\Mvariable}^6}}
   {9216}} \,,\nonumber
\end{eqnarray}
for $m=6$:
\begin{eqnarray}
&&a_{2}={\frac{-\Mvariable}{64}}\qquad
a_{3}={\frac{{{\Mvariable}^2}\,{\sqrt{\pi }}}{256}}
\nonumber\\
&&a_{4}=-\frac {\Mvariable^3}{384}\qquad
a_{5}=-\frac{{{\Mvariable}^2}\,
  \left( 15 - 4\,{{\Mvariable}^2} \right) \,{\sqrt{\pi }}}{8192}\label{m6} \\
&&a_{6}={\frac{-\left( {{\Mvariable}^3}\,
     \left( -20 + 3\,{{\Mvariable}^2} \right)  \right) }{11520}}\qquad
a_{7}={\frac{{{\Mvariable}^2}\,
   \left( 945 - 1160\,{{\Mvariable}^2} + 
    128\,{{\Mvariable}^4} \right) \,{\sqrt{\pi }}}{3145728}}
\nonumber
\end{eqnarray}
and for $m=7$:
\begin{eqnarray}
&&a_{2}={\frac{-\Mvariable}{120\,{\sqrt{\pi }}}}\qquad
a_{3}={\frac{{{\Mvariable}^2}}{480}}\nonumber\\
&&a_{4}={\frac{-{{\Mvariable}^3}}{720\,{\sqrt{\pi }}}}\qquad
a_{5}={\frac{{{\Mvariable}^2}\,
     \left( -6 + {{\Mvariable}^2} \right) }{3840}}\label{m7}\\
&&a_{6}={\frac{-\left( {{\Mvariable}^3}\,
       \left( -76 + 7\,{{\Mvariable}^2} \right)  \right) }{50400\,
     {\sqrt{\pi }}}}\qquad
a_{7}={\frac{{{\Mvariable}^2}\,
    \left( 27 - 15\,{{\Mvariable}^2} + {{\Mvariable}^4} \right) }
    {46080}} \,.\nonumber
\end{eqnarray}

\section{Calculation of the heat kernel asymptotics}
The information on the general structure of the heat kernel
asymptotic obtained in Sec. 2 can be summarized in the
following Lemma.
\par
\noindent {\bf Lemma 1.} {\it 1)
Let $N^\nu (f)=f_{;m\dots m}$ be the $\nu^{th}$ normal 
covariant derivative. There exist invariant local formulae
$a_{n,\nu }(y,D)$ so that:
\be
a_n(f,D)= \{ a_n(f,D_0;x)\} [M]
+\{ \sum_{0\le \nu\le n-1} N^\nu (f)
a_{n,\nu} (y,D) \} [\Sigma ]
\ee
\par\noindent 2)
If we expand $a_{n,\nu }$ with respect to a Weyl basis, the
coefficients depend on the dimension $m$ only through a
normalizing constant. 
\par\noindent 3)
Consider a transformation which changes the sign before the $m$-th
components of all vector and tensor fields and reverses the 
sign of the extrinsic curvature $L_{ab}$. Under this transformation
$a_{n,\nu} \to (-1)^\nu a_{n,\nu}$.
\par\noindent 4)
In the linear order of $V$
\be
a_n (f,D)=\{ V(z) \frac{\delta}{\delta E(z)} a_{n}(f,D_0)\}[\Sigma] \,.
\label{VE}
\ee
}
{\it Proof.} First assertion is now evident. 
Volume terms $ \{ a_n(f,D_0;x)\} [M]$
are given in the Appendix B.
One can also observe that
coefficients before monomials constructed from the geometric
invariants depend on the dimension $m$ only through the
factor $(4\pi )^{-m/2}$. A more simple way to prove the
second assertion is to consider product spaces $M=S^1\times M_1$
and $\Sigma =S^1\times \Sigma^1$ exactly repeating the corresponding
proof for manifolds with boundaries \cite{Gilkey}. Assertion 3)
follows from the fact that we can repeat all calculation of
section 2 with the replacement $x^m\to -x^m$. The last statement
of Lemma 1 is just a trivial generalization of the equation
(\ref{K1}) for non-unit smearing function $f$. Indeed, it is
sufficient to represent the variation on the r.h.s. of (\ref{VE}) as
\be
\frac{\delta}{\delta E(z)} K_0(t) =
\int_\M dx \int_0^t ds\, f(x)K_0(z,x;t-s)K_0(x,z;s) \,.
\ee
Assertion 4) is now evident.

Now we can determine several first heat kernel coefficient
up to a few yet undetermined constants.
\par\noindent {\bf Lemma 2.} {\it There exist universal constants
$c_{3,1},\dots ,c_{5,10}$ such that}
\begin{eqnarray}
a_0(f,D)&=&a_0(f,D_0) \nonumber \\
a_1(f,D)&=&0 \nonumber \\
a_2(f,D)&=&a_2(f,D_0)+ (4\pi )^{-m/2} \{ fV\}[\Sigma] \nonumber \\
a_3(f,D)&=&(4\pi )^{-(m-1)/2} \{c_{3,1} fV^2\}[\Sigma] \nonumber\\
a_4(f,D)&=&a_4(f,D_0)+ (4\pi )^{-m/2} \{ c_{4,1}fV^3 +\frac 1{6}
f\tau V +fEV \nonumber \\
& & +\frac 16 fV_{:aa} -\frac 16 f_{;m}VL_{aa} +\frac 16 f_{;mm}V
\}[\Sigma] \nonumber \\
a_5(f,D)&=& (4\pi )^{-(m-1)/2} \{c_{5,1} fV^4 +c_{5,2} f\tau V^2
+c_{5,3}\rho_{mm} V^2 \\
& & +c_{5,4} fV^2 E +c_{5,5} fV^2 L_{aa}L_{bb} 
+c_{5,6}fV^2 L_{ab}L_{ab} +c_{5,7}fV_{:aa}V \nonumber \\
& & +c_{5,8} fV_{:a}V_{:a} +c_{5,9}f_{;m} V^2L_{aa}
+c_{5,10}f_{;mm}V^2 \}[\Sigma ]
\nonumber
\end{eqnarray}
{\it Proof.} According to Lemma 1.1) and 2) any coefficient
$a_n$ contains all local invariants of appropriate dimension.
Some of the invariants, like e.g. $fVL_{aa}$, $f_{;m}V$ etc
are ruled out by Lemma 1.3). All terms linear in $V$ are
determined by Lemma 1.4).
\par\noindent
{\bf Lemma 3.} $c_{3,1}=\frac 18 ,\ c_{4,1}=\frac 16 ,\ 
c_{5,1}=\frac 1{64} ,\ c_{5,4}=\frac 18 ,\ c_{5,5}=-\frac 1{256} ,
\ c_{5,6}=\frac 1{128}$.
\par\noindent {\it Proof.} The coefficients $c_{3,1}$,
$c_{4,1}$, $c_{5,1}$, $c_{5,5}$ and $c_{5,6}$ are easily
calculated using the example of the delta-potential on the
sphere of the previous section. To calculate the coefficient
$c_{5,4}$ consider the case then $E=e{\bf 1}$ is a constant
proportional to the unit matrix. In this case $K(f;t)=
K(f;t)|_{E=0}\exp (te)$. This immediately gives 
$c_{5,4}=c_{3,1}=\frac 18$.

Several more universal constants can be calculated by a
reduction to Dirichlet and Neumann boundary value problems.
All necessary definitions and explicit expressions for
the heat kernel coefficient for that problems can be found
in the Appendix B.

\par\noindent {\bf Lemma 4.} {\it 1) Let $M=\Sigma\times [-a,a]$.
Let $\nabla_m =\partial_m$ and let all geometric invariants and
the smearing function $f$ be symmetric under $x^m\to -x^m$.
We suppose that $f$ and a sufficient number of its derivatives
vanish at $x^m=\pm a$. Then $a_n(f,D)=a_n(f,D_0,{\cal B}^-)+
a_n(f,D_0,{\cal B}^+)$ where the heat kernel coefficients
on the right hand side are calculated on $\Sigma\times [0,a]$,
and $S=\frac 12 V$.}
\par\noindent {\it Proof.}  Since reflection of the $m$th coordinate
commutes with $D$ one can subdivide the spectral resolution in the
two sets $(\lambda^N_\pm ,\phi^N_\pm )$ with normalized eigenfunctions
$\phi_\pm^N(-x^m)=\pm \phi_\pm^N(x^m)$. Then the heat kernel becomes
\be
K(f;t)=K_-(f;t)+K_+(f;t)
\ee
where
\begin{eqnarray}
&&K_\pm (f;t)=\int_M dx f(x) \sum_N \exp (-t\lambda_\pm )\phi_\pm (x)^2
\nonumber \\
&&\qquad =\int_0^a dx^m \int_\Sigma dz f(x^m,z) \sum_N 
\exp (-t\lambda_\pm )(\sqrt{2}\phi_\pm (x^m,z))^2 \,.
\end{eqnarray}
Now we observe that $\sqrt{2}\phi_\pm$ are normalized eigenfunctions
of the operator $D_0$ on $\Sigma\times [0,a]$ satisfying 
Dirichlet and Neumann  boundary conditions 
\be
\phi_-|_{x^m=0}=0 \qquad (\nabla_m +\frac 12 V)\phi_+|_{x^m=0}=0 \,.
\ee
Assertion of Lemma 4 follows immediately.

Under the conditions of Lemma 4 $L_{ab}$, $\rho_{mm}$ and $f_{;m}$
vanish identically. We can calculate only the coefficients of $f\tau V^2$,
$fV_{:aa}V$, $fV_{:a}V_{:a}$ and $f_{;mm}V^2$.
\par\noindent {\bf Corollary}
 $c_{5,2}=\frac 1{48}$, $c_{5,7}=\frac 1{24}$,
$c_{5,8}=\frac 5{192}$, $c_{5,10}=\frac 1{64}$

The rest of the universal constants can be calculated by
using the conformal properties of the heat kernel which are
exactly the same as for the usual boundary value problem
\cite{bg90}.
\par\noindent
{\bf Lemma 5.} {\it If $D(\epsilon )=e^{-2f\epsilon}D$, then
$\frac d{d\epsilon}|_{\epsilon =0}a_n(1,D)=(m-n)a_n(f,D)$.}

Under the conformal transformation the metric $g$ acquires
a multiplier $e^{2f\epsilon}$. $V$ is transformed to $e^{-f\epsilon}V$.
Basic geometric quantities transform as \cite{bg90}
\begin{eqnarray}
(\frac d{d\epsilon} |_{\epsilon =0} \Gamma )_{ij}^k
&=& \delta  \sb{
ik}f \sb{ ;j}+ \delta  \sb{ jk}f \sb{ ;i}- \delta  \sb{
ij}f \sb{ ;k} \nonumber \\
 (\frac d{d\epsilon} |_{\epsilon =0} L )_{ab}
&=& - \delta  \sb{ ab}f \sb{
;m}-fL \sb{ ab} \nonumber \\
\frac d{d\epsilon} |_{\epsilon =0} (E)&=& -2fE +\frac 12 (m-2)
f_{;ii} \nonumber \\
\frac d{d\epsilon} |_{\epsilon =0} (\tau )&=& -2f\tau +2(1-m)
f_{;ii} \nonumber \\
\frac d{d\epsilon} |_{\epsilon =0} (\rho_{mm})&=&
-2f\rho_{mm}-f_{;aa} +(1-m)f_{;mm} \,,\label{confbas}
\end{eqnarray}
where $\Gamma$ is the Christoffel connection. We need the following
conformal relations
\begin{eqnarray}
\frac d{d\epsilon} |_{\epsilon =0} V^2\tau &=&
-4fV^2\tau +2(1-m)[f_{;mm}V^2 +2f(V_{:aa}V+V_{:a}V_{:a})
 \nonumber \\
& & -L_{:aa}f_{;m} V^2] \nonumber \\
\frac d{d\epsilon} |_{\epsilon =0} V^2\rho_{mm} &=&
-4V^2\rho_{mm} f -2f(V_{:aa}V+V_{:a}V_{:a})+L_{aa}Vf_{;m} 
\nonumber \\
& & +(1-m) V^2f_{;mm} \nonumber \\
\frac d{d\epsilon} |_{\epsilon =0} V^2 E &=&
-4V^2Ef+(m-2)f(V_{:aa}V+V_{:a}V_{:a}) \nonumber \\
& & -\frac 12 (m-2)L_{aa}f_{;m}V^2 
+\frac 12 (m-2) f_{;mm}V^2 \nonumber \\
\frac d{d\epsilon} |_{\epsilon =0} V^2 L_{aa}^2 &=&
-4V^2 L_{aa}^2 f -2 (m-1)V^2 L_{aa}f_{;m} \nonumber \\
\frac d{d\epsilon} |_{\epsilon =0} V^2 L_{ab}L_{ab} &=&
-4V^2 L_{ab}L_{ab}f-2V^2 L_{aa}f_{;m} \nonumber \\
\frac d{d\epsilon} |_{\epsilon =0} V_{:aa}V &=&
-4V_{:aa}Vf -(m-3)(V_{:aa}Vf+V_{:a}V_{:a}f) \nonumber \\
\frac d{d\epsilon} |_{\epsilon =0} V_{:a}V_{:a} &=&
-4V_{:a}V_{:a}f +2(V_{:aa}Vf+V_{:a}V_{:a}f) \,. \label{confmon}
\end{eqnarray}
To obtain the relation (\ref{confmon}) we used integration by
parts. Let $n=5$. By collecting the terms with $V_{:aa}Vf$
and $V^2L_{:aa}f_{;m}$ we obtain
\begin{eqnarray}
&&0=4(1-m)c_{5,2}-2c_{5,3}+(m-2)c_{5,4}-(m-3)c_{5,7}
+2c_{5,8} \nonumber \\
&&0=-2(1-m)c_{5,2}+c_{5,3}-\frac 12 (m-2)c_{5,4}
-2(m-1)c_{5,5} -2c_{5,6} -(m-5)c_{5,9} \,.\nonumber
\end{eqnarray}
Solving these equation one obtains $c_{5,3}=1/192$ and 
$c_{5,9}=-5/384$. Below we collect for convenience
of the reader all universal constants of Lemma 2
\begin{eqnarray}
&& c_{3,1}=\frac 18 ,\ c_{4,1}=\frac 16 ,\ 
c_{5,1}=\frac 1{64} ,\ c_{5,2}=\frac 1{48} ,\ 
c_{5,3}=\frac{1}{192} ,\ c_{5,4}=\frac 18 ,
\label{allconst} \\
&&\ c_{5,5}=-\frac 1{256} ,
\ c_{5,6}=\frac 1{128} ,\   
c_{5,7}=\frac 1{24} , \ 
c_{5,8}=\frac 5{192} , \
c_{5,9}=-\frac{5}{384} ,\ 
c_{5,10}=\frac 1{64} \,.
\nonumber
\end{eqnarray}

\section{Conclusions}
In the present paper we have studied the heat kernel expansion
for for an ope\-rator of Laplace type in the presence of
semitransparent boundaries. We have determined the general
form of the asymptotic expansion. Namely, we proved the validity
of the asymptotic series (\ref{asymptotex}). We have calculated
explicitly several first terms of the expansion for the most
general operator of Laplace type and arbitrary boundary potential.
We believe that this is the most complete study done
in this field so far.

Our methods of deriving the heat kernel coefficients
admit extensive cross-checking. Most of the universal
constants can be calculated by at least two independent
methods. If needed, one can calculate the higher coefficients
as well. As possible generalizations we can suggest
the $\delta'$ potentials or even general four-parameter
family on matching conditions \cite{Albeverio} on the hypersurface
$\Sigma$. Another possible development of the present results
can be the renormalization of quantum field theory in the
presence of singular interactions \cite{Symanzik}. We believe
that semitransparent boundaries provide a more adequate framework
for the renormalization than ``abrupt'' boundary conditions
of Dirichlet or Neumann type. For most recent work on renormalization
with singular potentials see \cite{Solodukhin}.

\section*{Acknowledgments}
We are grateful to Klaus Kirsten for fruitful discussions.
One of the authors (D.V.) thanks the Alexander von Humboldt
foundation and RFBR, grant 97-01-01186. He is also grateful
to S. Randjbar-Daemi for kind hospitality at the Abdus Salam
ICTP where this work was completed.

\section*{Appendix A}
Some first non zero coefficients $X_{n,i}$ appearing in
\Ref{expansion} read:
 $$   X_{1,1}={\frac{\Mvariable}{2}}, ~ 
X_{2,2}={\frac{-{{\Mvariable}^2}}{8}}, ~ 
X_{3,3}={\frac{\Mvariable}{16}} + {\frac{{{\Mvariable}^3}}{24}}, ~ 
X_{3,5}={\frac{-3\,\Mvariable}{8}}, ~ 
X_{3,7}={\frac{5\,\Mvariable}{16}}, ~ $$
$$
X_{4,4}={\frac{-{{\Mvariable}^2}}{32}} - {\frac{{{\Mvariable}^4}}{64}}, ~ 
X_{4,6}={\frac{3\,{{\Mvariable}^2}}{16}}, ~ 
X_{4,8}={\frac{-5\,{{\Mvariable}^2}}{32}}, ~ $$
$$
X_{5,5}={\frac{27\,\Mvariable}{256}} + {\frac{{{\Mvariable}^3}}{64}} +
     {\frac{{{\Mvariable}^5}}{160}}, ~
     X_{5,7}={\frac{-145\,\Mvariable}{64}} -
     {\frac{3\,{{\Mvariable}^3}}{32}}, ~
     X_{5,9}={\frac{1085\,\Mvariable}{128}} +
     {\frac{5\,{{\Mvariable}^3}}{64}}, ~ $$
$$
   X_{5,11}={\frac{-693\,\Mvariable}{64}}, ~ 
X_{5,13}={\frac{1155\,\Mvariable}{256}}. ~ 
$$
The Zeta functions defined in Eq. \Ref{sum} are
\beao \zeta(2,n)&=&2\zeta_{\rm R}(2s+n)\\ \zeta(3,n)&=&2\zeta_{\rm
H}(2s+n-1;\frac12)\\ \zeta(4,n)&=& \zeta_{{\rm R}}(2s+n-2)\\
\zeta(5,n)&=&{2\over 3!}(\zeta_{\rm
H}(2s+n-3;\frac12)-\frac14\zeta_{\rm H}(2s+n-1;\frac12))\\
\zeta(6,n)&=& {2\over 4!}(\zeta_{{\rm R}}(2s+n-4)-\zeta_{{\rm
R}}(2s+n-2))\\ \zeta(7,n)&=& {2\over 5!}(\zeta_{\rm
H}(2s+n-5;\frac12)-\frac52 \zeta_{\rm
H}(2s+n-3;\frac12)+\frac{9}{16}\zeta_{\rm H}(2s+n-1;\frac12)) \,,
\eeao
where $\zeta_{\rm R}$ and $\zeta_{\rm H}$ are the Riemann and Hurwitz zeta
functions correspondingly.

\section*{Appendix B}
In this appendix we give expressions for the heat kernel
coefficients for the Dirichlet and Neumann boundary value
problems. Let $M$ be a smooth compact Riemannian manifold
with smooth boundary $\partial M$. Let $S$ be an endomorphism
on ${\cal V}_{\partial M}$ and let $\phi_{;m}$ be a
covariant derivative of $\phi$ with respect to inward unit
normal. We define the modified Neumann boundary operator
${\cal B}^+$ and the Dirichlet boundary operator ${\cal B}^-$
by
\be
{\cal B}^+\phi :=(\phi_{;m}+S\phi )|_{\partial M} \qquad
{\cal B}^-\phi := \phi |_{\partial M}
\label{boundop}
\ee
We set $S=0$ for the Dirichlet boundary conditions to have
a uniform notations.

 We need only the case of totally geodesic boundary
($L_{ab}=0$). We drop certain boundary invariants which vanish under
conditions of Lemma 4. Several first heat kernel coefficients
are \cite{bg90,a5bgv,a5kirsten}
\begin{eqnarray}
a_0(f,D,{\cal B}^\pm )&=& (4 \pi ) \sp{ -m/2} {\rm Tr} (f)[M] \nonumber \\
a_1(f,D,{\cal B}^\pm )&=& \pm \frac 14 (4 \pi ) \sp{
  -(m-1)/2} {\rm Tr} (f)[ \partial M] \nonumber \\
a_2(f,D,{\cal B}^\pm )&=& (4 \pi ) \sp{ -m/2} 
\frac 16 {\rm Tr}\{ (6FE+F\tau ) [M] +12fS[\partial M] \}
\nonumber \\
a_3(f,D,{\cal B}^\pm )&=& \pm \frac 1{384}
(4 \pi ) \sp{ -(m-1)/2} {\rm Tr} \{ f(96E+16 \tau 
+192S \sp{ 2})
+24F \sb{ ;mm} \} [\partial M] \nonumber \\
a_4(f,D,{\cal B}^\pm )&=& (4 \pi ) \sp{ -m/2} \frac 1{360} {\rm Tr}
(f(60E \sb{
  ;kk}+60 \tau E+180E \sp{ 2}+30 \Omega  \sp{2}
  +12 \tau  \sb{ ;kk}
\nonumber \\
& &+5 \tau  \sp{ 2}-2 \rho  \sp{
  2}+2R \sp{ 2})[M] +(f(720SE +120S\tau +480S^3 
\label{DirNeu} \\
& & +120S_{:aa}) + 120f_{;mm}S) [\partial M] ) \nonumber \\
a_5(f,D,{\cal B}^\pm )&=&\pm \frac 1{5760} (4\pi )^{-(m-1)/2}
{\rm Tr} \{ f (360E \sb{ ;mm}+1440E \sb{ ;m}S+720E \sp{
2}
\nonumber \\
& &+2880ES \sp{ 2} 
+1440S \sp{ 4 }
 +240E \sb{ :aa}+240 \tau E+120 \Omega  \sb{
ab} \Omega  \sb{ ab} \nonumber \\
& &+20 \tau  \sp{
2}-8 \rho  \sp{ 2}+8R \sp{ 2 } + 480 \tau S^2 +960 S_{:aa}S
+600 S_{:a}S_{:a}) \nonumber \\
& & +f_{;mm} (360E+360S \sp{ 2}+60 \tau ) +45f_{;mmmm} \}[\partial M] \,.
\nonumber 
\end{eqnarray}
On manifold without a boundary one should keep volume contributions
only.

\bibliographystyle{sort}

\end{document}